\documentclass[12pt]{article}
\usepackage{epsfig,amsfonts,amssymb}
\usepackage{hyperref}
\usepackage{upgreek}
\pdfoutput=1

\usepackage[margin=2.3cm]{geometry}
\usepackage{amsmath,amssymb,amsthm,bm}
\usepackage{hyperref}
\usepackage[dvipsnames]{xcolor}

\usepackage{amsmath,amssymb,bm,booktabs,array}

\usepackage[dvipsnames]{xcolor}

\usepackage{cite}
\usepackage{cite}

\usepackage{comment}

\def\ZZZ{{\hbox{ Z\kern-1.6mm Z}}}
\def\RRR{{\hbox{ R\kern-2.4mm R}}}
\def\CCC{{\hbox{ C\kern-2.0mm C}}}
\def\zzz{{\hbox{z\kern-1mm z}}}

\def\ZZZ{\mathbb{Z}}
\def\RRR{\mathbb{R}}

\newcommand{\bt}{\uptau}
\newcommand{\bbt}{\bar\uptau}
\newcommand{\bxi}{\xi}
\newcommand{\xit}{\widetilde\xi}
\newcommand{\at}{\widetilde{a}}
\newcommand{\alt}{\widetilde{\alpha}}

\newcommand{\bq}{\bm{q}}
\newcommand{\bqb}{\bar{\bm{q}}}
\newcommand{\by}{\bm{y}}
\newcommand{\byt}{\widetilde{\bm{y}}}

\newcommand{\vt}{\vartheta}

\newcommand{\qeq}{{\hbox{=\kern-2.3mm ? \kern.5mm }}}
\renewcommand{\qeq}{=}

\newcommand{\eps}{\epsilon}

\newcommand{\vp}{\varphi}

\newcommand{\FF}{{\cal F}}

\newcommand{\MM}{{\cal M}}

\newcommand{\XX}{{\cal X}}

\newcommand{\wt}{\widetilde}
\newcommand{\wh}{\widehat}

\newcommand{\SSS}{{\cal S}}

\newcommand{\be}{\begin{equation}}
\newcommand{\ee}{\end{equation}}
\newcommand{\ben}{\begin{eqnarray}\displaystyle}
\newcommand{\een}{\end{eqnarray}}

\newcommand{\refb}[1]{(\ref{#1})}
\newcommand{\p}{\partial}
\newcommand{\sectiono}[1]{\section{#1}\setcounter{equation}{0}}

\def\one{{\hbox{ 1\kern-.8mm l}}}
\def\zero{{\hbox{ 0\kern-1.5mm 0}}}

\newcommand{\bea}[1]{\begin{eqnarray}\label{#1} }
\newcommand{\eea}{\end{eqnarray}}

\newcommand{\ZB}{{\bf Z}}

\begin{document}

\baselineskip 24pt

\begin{center}

%{\Large \bf How to Study Isolated Exotic Branes}
{\Large \bf Revisiting String Theory on $AdS_3\times S^3\times K3$}

\end{center}

\vskip .6cm
\medskip

\vspace*{4.0ex}

\baselineskip=18pt

\centerline{\large \rm Ashoke Sen}

\vspace*{4.0ex}

\centerline{\large \it International Centre for Theoretical Sciences - TIFR 
}
\centerline{\large \it  Bengaluru - 560089, India}

%\centerline{\large \it ~$^c$Homi Bhabha National Institute}
%\centerline{\large \it Training School Complex, Anushakti Nagar,
%    Mumbai 400085, India}

\vspace*{1.0ex}
\centerline{\small E-mail:  ashoke.sen@icts.res.in}

\vspace*{5.0ex}

\centerline{\bf Abstract} \bigskip

Recently Eberhardt, Gaberdiel, Murthy and Rangamani found that the world-sheet 
computation
of the partition function of string theory on $AdS_3\times S^3\times K3$ at finite temperature and
chemical potential
suffers from an ambiguity since the superghost contribution to the world-sheet partition function
blows up and the matter fermion contribution vanishes. They fixed this ambiguity by requiring the 
results to be consistent with space-time supersymmetry.
We show that the derivatives of the
logarithm of the partition function with respect to the temperature and chemical potentials 
are unambiguous and can be computed without having to require space-time supersymmetry.
From this we can recover the logarithm of the partition function up to an additive constant.
The results agree with those of 
Eberhardt et al. The correct choice of the picture changing operator locations plays a
crucial role in this analysis.

\vfill \eject

\tableofcontents

\sectiono{Introduction} \label{s1}

String theory on $AdS_3\times S^3\times \MM$, with $\MM=K3$ or $T^4$, 
has been of interest since the discovery that it is dual to a two dimensional super-conformal
field theory\cite{9711200}. 
The original background had RR 3-form flux through $AdS_3$ and $S^3$,
but if we replace these by NSNS flux then the string theory has 
a world-sheet description based on the standard
RNS formalism\cite{9806194,9903219} 
and explicit computations beyond supergravity approximation 
become possible on the string theory side\cite{0001053,0005183,0111180}.
In particular the 
partition function of this theory
at finite temperature and chemical potential
is expected to be dual to the partition function of
the dual SCFT. 
Further progress on this was made recently with the realization that in appropriate limit,
the dual CFT is a symmetric product orbifold and hence is also 
solvable\cite{1803.04423,1812.01007,1911.00378,2009.11306,2011.10038,2306.16280}.
Hence one can compute physical quantities on both sides of the conjectured duality 
and prove the duality.

In a recent paper\cite{2610.20696}, Eberhardt, Gaberdiel, Murthy and Rangamani (EGMR)
explicitly computed the  partition function of string theory on 
$AdS_3\times S^3\times K3$ by expressing it as the exponential of the
moduli space integral of the partition
function $Z_{ws}$ of the world-sheet theory and found that the result is ambiguous. 
The signs of such ambiguities were already seen in the earlier 
analysis of \cite{2408.14567,2505.14380}.
The ambiguity comes
from the fact that the partition function of the superconformal $\beta$-$\gamma$ ghosts diverges
when we impose periodic boundary condition on these fields, and in the same sector the
partition function of the matter fermion zero modes vanishes.
EGMR resolved this problem by requiring the result to be consistent with space-time 
supersymmetry.

While this resolves the problem for this particular background, this
raises a serious question on world-sheet string theory. World-sheet theory (possibly augmented
by string field theory to deal with infrared issues) is supposed to give
a complete description of perturbative string theory. Space-time supersymmetry is supposed
to come out of the theory and should not be an input. Furthermore, if we really need
additional inputs of this type, then it would be no longer clear what string theory means when 
we have a general background which may not have space-time supersymmetry. For this reason
it is important to resolve the issue as to whether or not world-sheet string theory by itself can
compute the partition function of string theory on $AdS_3\times S^3\times K3$.

With this motivation we now review the origin of the problem. In the world-sheet description
of string theory, it is customary to use $-1/2$ and $-3/2$ picture states to describe the
Ramond sector. We need both these pictures since they are BPZ conjugates of each other.
Let us define
\be
|q\rangle = e^{q\phi}(0)|0\rangle
\ee
to be the 'vacuum' of picture number $q$, although these are not the lowest $L_0$
eigenstates in the corresponding picture numbers. Here $\phi$ is a world-sheet scalar field,
arising out of
bosonization of the $\beta$-$\gamma$ ghost system:
\be \label{e1.2}
\beta = \p\xi \, e^{-\phi}, \qquad \gamma=\eta\, e^\phi\, ,
\ee
where $\xi$ and $\eta$ are dimension (0,0) and (0,1) holomorphic fields respectively. Then
in the Ramond (R) sector the world-sheet SCFT has 
infinite number of degenerate states
of the form:
\be
\beta_0^n |-3/2\rangle, \qquad \gamma_0^n|-1/2\rangle\, ,
\ee
for any non-negative integer $n$.
Furthermore, up to an overall normalization these states are BPZ conjugates of each other.
As a result, in the trace the sum over these states will give infinite contribution since
they all have the same $L_0$ eigenvalues. This is essentially the reason for the divergence
in the partition function of the $\beta$-$\gamma$ system.

In the world-sheet path integral formulation the divergence is seen as follows. First of all,
note that the $\gamma_0$ and $\beta_0$ are GSO odd, and hence only alternate values
of $n$ appear in the trace. In the path integral formulation this is implemented by summing
over periodic and anti-periodic boundary conditions along the b-cycle of the world-sheet
torus. The anti-periodic boundary condition on Ramond sector states produces a factor of $(-1)^n$
in the trace and the divergence described above cancels while the periodic boundary condition
weighs all states equally and produces the divergence. Hence the divergence arises when
we have periodic boundary conditions along both cycles of the world-sheet torus. In 
the Euclidean path integral formalism, 
it is precisely with these boundary conditions that the constant modes of $\beta$ and
$\gamma$ become zero modes and produce divergence.
A similar phenomenon is present
in the anti-holomorphic sector.  This is exactly what is observed in the analysis of
\cite{2610.20696}.

This problem is present in all string backgrounds, but it does not affect the string
amplitude for the following reason. This is easiest to describe in the language of string field
theory where the Ramond sector propagator contains an insertion of zero mode $\XX_0$ of the
picture changing operator\cite{1501.00988,1508.05387}. 
Since $\XX_0$ carries picture number one, its matrix element vanishes
between a picture number $-1/2$ and a picture number $-3/2$ state. Instead it has
non-zero matrix element between a pair of picture number $-3/2$ states, and it is precisely these
kinds of matrix elements that appear in string field theory amplitude computation. Hence if instead
of the world-sheet partition function
we had been computing a torus one point function, 
then the contribution from the R sector internal propagator will 
involve a matrix element of the form
\be
\langle a | \XX_0| b\rangle
\ee
where $|a\rangle$ and $|b\rangle$ are both $-3/2$ picture states. 
The infinite degeneracy is still present among these states but since $\gamma_0$ annihilates
$-3/2$ picture vacuum, for both $|a\rangle$ and $|b\rangle$, the degeneracy arises due to the
action of $\beta_0$. Since the action of $\beta_0$ lowers the ghost number by 1, we see that
the matrix element will vanish by ghost number conservation except possibly for a finite number
of states in a given level\cite{1703.06410}.

This suggests that if instead of the string partition function $\ZB$,
we compute the derivative of 
$\ln\ZB$ with respect to the temperature and other chemical potentials, 
and express this as the moduli space integral of the torus one
point function of appropriate closed string vertex operators, then there should be no infinities
from the infinite degeneracy of states in the $\beta$-$\gamma$ sector, and we should
get an unambiguous result. This is the strategy
we are going to follow.
Once we have computed the derivatives of  $\ln\ZB$, we can recover $\ln\ZB$ by integrating
these equations up to an additive constant.

We note in passing that this is very similar to the strategy that was used in computing the
partition function of type II string theory in a background with non-zero Euler number. The
naive result is divergent, but the derivative of the partition function with respect to the
dilaton gives an unambiguous answer once we make the correct choice of PCO
positions\cite{2607.12031}. 

Our main results are given in \refb{e3.21yx}, \refb{e3.41x},
\refb{e3.47x} and
\refb{e3.52x} which we reproduce here:
\be \label{e3.21yxint}
\delta  f^{01}_{ee}  =
\bigg[
\delta \uptau \, {\p\over \p\uptau} + 
\delta \bar \uptau  \, {\p\over \p\bar\uptau} + \delta \xi \, {\p\over \p\xi} \,  + 
\delta \wt\xi   \, {\p\over \p\wt\xi}
\bigg] \bigg[ \uptau_2
 \int_\SSS {d^2\tau\over \tau_2}
\,  \wh Z_{ee} \bigg]\, ,
\ee
\be \label{e3.41xint}
\delta  f^{01}_{oo} = - {1\over \pi \, i\, k} 
 \int_\SSS d^2\tau\, 
  \bigg[
\delta \uptau \, {\p^2\wh Z_{oo}\over \p\xi\p\wt\xi} 
-
\delta \bar \uptau  \, {\p^2\wh Z_{oo}\over \p\xi\p\wt\xi}  +\delta \xi \, 
{\p^2\wh Z_{oo}\over \p\uptau\p\wt\xi} 
 -
\delta \wt\xi   \, 
{\p^2\wh Z_{oo}\over \p\bar\uptau\p\xi} 
\bigg]\, ,
\ee
\be \label{e3.47xint}
\delta f^{01}_{eo} =  {1\over i\,  \pi}\int_\SSS d^2\tau\, 
  \bigg[
\delta \uptau \, {i\pi\uptau_2\over \tau_2} {\p \wh Z_{eo}\over \p \xi}+{1\over k} 
\delta \bar \uptau  \,   {\p^2 \wh Z_{eo}\over \p\xi \p \bar\uptau}
+\delta \xi \, \, {i\pi\over \tau_2} {\p \over \p \uptau} (\uptau_2 \wh Z_{eo})
+{1\over k}
\delta \wt\xi    \,  {\p^2 \wh Z_{eo}\over \p\xi \p \wt\xi}
\bigg]\, ,
\ee
\be \label{e3.52xint}
\delta f^{01}_{oe} =  {1\over i\, \pi}\int_\SSS d^2\tau\, 
  \bigg[-\delta \uptau  \,  {1\over k}  {\p^2 \wh Z_{oe}\over \p\wt\xi \p \uptau} +
\delta \bar\uptau \, {i\pi\uptau_2\over \tau_2} {\p \wh Z_{oe}\over \p \wt\xi} 
- {1\over k}\, 
\delta \xi    \,   {\p^2 \wh Z_{oe}\over \p\wt\xi \p \xi}
+ \delta \wt\xi \, {i\pi\over \tau_2} {\p \over \p \bar\uptau}
(\uptau_2  \wh Z_{oe})\bigg]\, .
\ee
Here $f^{01}$ is the correct partition function in the winding number one sector and
$\uptau_2 \int d^2\tau \tau_2^{-1} \wh Z$ is the `naive'  partition function in the same sector.
$\uptau=\uptau_1+i\uptau_2$ and $\bar\uptau$ are the chemical potentials associated with the 
isometries of $AdS_3$ and $\xi$ and $\wt\xi$
are the chemical potentials associated with the isometries of $S^3$.
$\SSS$ is the strip in the complex $\tau=\tau_1+i\tau_2$ plane spanning the range
$\tau_2>0$, $|\tau_1|<1/2$. The subscripts $o$ and $e$ denote odd and even spin
structures, both in the anti-holomorphic (left) and holomorphic (right) sectors. $\delta$
denotes the change under arbitrary variation of $\uptau$, $\bar\uptau$, $\xi$, $\wt\xi$.
In particular \refb{e3.21yxint} shows that in the even - even spin structure sector, 
we have the expected
result
\be \label{e3.21yxintnew}
f^{01}_{ee}  =
\bigg[ \uptau_2
 \int_\SSS {d^2\tau\over \tau_2}
\,  \wh Z_{ee} \bigg]\, ,
\ee
up to an undetermined additive constant.

The rest of the paper is organized as follows. In section \ref{s2} we review the
world-sheet results and the problem that we face in computing the thermal partition function.
In section \ref{s3} we compute the derivative of the thermal partition function with respect to
the temperature and various chemical potentials. In section \ref{s4} we show that the
proposed expression for the partition function based on space-time 
supersymmetry\cite{2610.20696} agrees
with our results. In appendix \ref{sb} we define the analog of the world-sheet field $X^\mu$ in
flat target space for $AdS_3$ target space. This field has been used in the analysis leading
to \refb{e3.21yxint}-\refb{e3.52xint}.
Appendix \ref{sa} contains the derivation of a technical result involving 
two point function of the derivative of this field with various currents on the world-sheet torus.

\sectiono{The problem} \label{s2}

In this section we shall describe in more detail the problem that we encounter in
computing the  partition function of string theory on $AdS_3\times S^3\times K3$. In this
process we shall also collect all the world-sheet results that we shall need for our analysis.

\subsection{World-sheet results}

The target space metric and two form fields on Euclidean $AdS_3\times S^3\times K3$ 
are given by
\ben\label{e2.1}
&& ds^2 = k \left\{ d\rho^2 + \cosh^2\rho\, dt^2 + \sinh^2\rho \, d\vp^2\right\}
+ k \left\{ d\theta^2 + \cos^2\theta \, d\phi_1^2 + \sin^2\theta \, d\phi_2^2\right\} + ds_{K3}^2\, ,
\nonumber \\
&& {1\over 2} B_{\mu\nu} dx^\mu\wedge dx^\nu = i\, k\sinh^2\rho \, dt\wedge d\vp
+ k \sin^2\theta \, d\phi_1\wedge d\phi_2\, .
\een
Here $\rho,t,\vp$ are the coordinates of $AdS_3$, or equivalently the $SL(2,R)$ group
manifold and $\theta,\phi_1$ and $\phi_2$ are the coordinates of $S^3$, or equivalently the
$SU(2)$ group manifold.  The covering space of the manifold \refb{e2.1}
has $SL(2,R)_L\times SL(2,R)_R\times SU(2)_L\times SU(2)_R$ isometry associated with the $AdS_3\times S^3$ factor. $ds_{K3}^2$ is the metric on $K3$.
The full world-sheet theory with this target space
consists of the following components:
\begin{enumerate}
\item An $SL(2,R)$ WZW model of level $k$ associated with
the $AdS_3$ factor.
\item Three holomorphic free fermions $\psi^1,\psi^2,\psi^3$ and three
anti-holomorphic  free fermions $\bar\psi^1,\bar\psi^2,\bar\psi^3$, that form
the superpartners of the bosonic currents of the $SL(2,R)$ WZW model.
\item
An $SU(2)$ WZW model of level $k$ associated with the $S^3$ factor.
\item
Three holomorphic free fermions $\chi^1,\chi^2,\chi^3$ and three
anti-holomorphic  free fermions $\bar\chi^1,\bar\chi^2,\bar\chi^3$, that form
the superpartners of the bosonic currents of the $SU(2)$ WZW model.
\item A superconformal field theory associated with the world-sheet sigma model
with target space $K3$.
\item The usual $b,c,\beta,\gamma$ ghost system and their anti-holomorphic 
counterparts. For these fields we shall use the same normalization conditions as
in \cite{2405.19421}. 
\end{enumerate}

We denote by $\bar J^3, J^3,\bar K^3, K^3$ the currents associated with the appropriate
U(1) subgroups of $SL(2,R)_L$,  $SL(2,R)_R$, $SU(2)_L$ and $SU(2)_R$ respectively.
Their linear combinations generate translations along the $t,\vp,\phi_1,\phi_2$ directions  and
 also generate appropriate rotations in the $\psi^1$-$\psi^2$,
$\bar\psi^1$-$\bar\psi^2$, $\chi^1$-$\chi^2$,
$\bar\chi^1$-$\bar\chi^2$ planes.
The singular parts of the operator product expansion of these currents take 
the form:\footnote{To simplify notation, we shall write the argument of a vertex operator as a single complex variable,
{\it e.g.} $z$, even though in general it depends on $z$ and $\bar z$ and sometimes can
depend only on $\bar z$. This is the case for example for $\bar J^3(z)$.}
\ben
&& \label{e2.4}
J^3(z) \, J^3(w) =-{k\over 2 (z-w)^2}, \qquad K^3(z) \, K^3(w) ={k\over 2 (z-w)^2}, \nonumber \\
&& \bar J^3(z) \, \bar J^3(w) =-{k\over 2 (\bar z-\bar w)^2},\qquad 
\bar K^3(z) \, \bar K^3(w) ={k\over 2 (\bar z-\bar w)^2}\, .
\een 
We also have
\ben \label{e2.5}
&& \psi^3(z) \psi^3(w) =  -{k\over 2 (z-w)}, \qquad \chi^3(z) \chi^3(w) =  {k\over 2 (z-w)},
\nonumber \\
&& \bar\psi^3(z) \bar\psi^3(w) =  -{k\over 2 (\bar z-\bar w)}, \qquad
\bar\chi^3(z) \bar\chi^3(w) =  {k\over 2 (\bar z-\bar w)}\, .
\een
The relevant operator product expansion 
in the superconformal ghost sectors are:
\be
e^{q_1\phi}(z) e^{q_2\phi}(w) 
= (z-w)^{-q_1q_2} e^{(q_1+q_2)\phi}(w), \quad \p\xi(z) \eta(w) = -{1\over (z-w)^2}\, ,
\ee
where $\xi,\eta,\phi$ are related to the $\beta,\gamma$ system via \refb{e1.2}.

The superconformal currents are given by\footnote{These expressions are correct for the
bosonic part of the currents $J^a$ and $K^a$ but the coefficient is different for the fermionic
contribution to the currents. The correct normalization is determined from the operator product expansions
\refb{e3.17x}, \refb{e3.18x}.
}
\ben \label{e3.11a}
T_F(z) &=& {1\over k} \psi^a (z) J^a(z)+{1\over k} \chi^a(z) K^a(z)+T_F^{K3}
+ T_F^{ghost}, 
\nonumber \\ \bar T_F(z) &=& {1\over 
k} \bar\psi^a (z) \bar J^a(z) +{1\over k} \bar\chi^a(z) \bar K^a(z)+\bar T_F^{K3}
+ \bar T_F^{ghost}\, .
\een
The holomorphic picture changing operator $\XX$ is given by
\be
\XX = c\p\xi + e^\phi T_F -{1\over 4} \p\eta \, e^{2\phi}\, b -{1\over 4} \p \left( \eta \, e^{2\phi}\, b
\right)\, .
\ee
Similar definition holds for the anti-holomorphic sector.

We have the relations
\be\label{e3.8}
J^3 = -G_{-1/2}\psi^3, \qquad \bar J^3 = -\bar G_{-1/2} \bar\psi^3, \qquad 
K^3 = G_{-1/2}\chi^3, \qquad \bar K^3 = \bar G_{-1/2} \bar\chi^3\, ,
\ee
where
\be
G_n = 2\, \ointop dz \, z^{n+1/2} T_F(z), \qquad \bar G_n = 2\, 
\ointop d\bar z \, \bar z^{n+1/2} \bar
T_F(\bar z)\, .
\ee
$\ointop$ is normalized so that 
\be\label{enormint}
\ointop dz / z=1\, , \qquad \ointop d\bar z / \bar z=1\, .
\ee

One can easily check using \refb{e2.1}
that $t$ is a harmonic function. Furthermore, it is single valued before we
make the periodic identification under translation by $2\pi\uptau_2$. 
Hence it represents a dimension 
(0,0) primary and we have\footnote{We shall denote the world-sheet field associated
with the target space-coordinate $t$ by the same symbol $t$.}
\be\label{e3.9}
\p t = L_{-1} t = (G_{-1/2})^2  t, \qquad \bar\p t = \bar L_{-1} t= (\bar G_{-1/2})^2  t\, .
\ee
$t$ is the analog of the world-sheet field $X$ in flat target space-time. In appendix \ref{sb}
we have defined $t$ in terms of the primary fields of the $SL(2,R)$ current algebra. 
$i\, J^3_0$ and $i\, \bar J^3_0$ each generates
1/2 unit of translation along $t$ besides generating translation along $\vp$.
This gives
\be
i\ointop dz J^3(z)\, t (0)=- i/2, \qquad i\ointop d\bar z \bar J^3(z) \, t(0) =- i/2\, ,
\ee
and
\be \label{e3.14e}
J^3(z)\, t (w) = -{1\over 2 (z-w)}, \qquad  \bar J^3(z) \, t(w) = -{1\over 2 (\bar z -\bar w)}\, ,
\ee
up to non-singular terms.
Let us denote by $f_1(\rho,t,\vp)$ and $f_2(\rho,t,\vp)$ the action of $J^1$ and $J^2$ on $t$, 
and by $\bar f_1(\rho,t,\vp)$ and $\bar f_2(\rho,t,\vp)$ the action of $\bar J^1$ and 
$\bar J^2$ on $t$.
Then we get
\be \label{e3.11}
\psi^t\equiv G_{-1/2} t = -{1\over k} \left\{
\psi^3 -2\, \psi^1 f_1 -2\, \psi^2 f_2\right\}, \qquad \bar\psi^t\equiv \bar G_{-1/2} t =- {1\over  k}
\left\{\bar\psi^3
-2\, \bar\psi^1 \bar f_1 -2\, \bar\psi^2 \bar f_2\right\}\, .
\ee
\refb{e3.9} now gives
\be
G_{-1/2} \psi^t = \p t, \qquad \bar G_{-1/2} \bar\psi^t =\bar\p t\, .
\ee
For later use, we shall write down the leading terms in the operator product expansion of
$T_F$ and $\bar T_F$ with various other fields that follow from this analysis:
\be \label{e3.17x}
T_F(z) \psi^3(w) =- {J^3(w)\over 2(z-w)} , \quad T_F(z) \chi^3(w) = {K^3(w)\over 2(z-w)} ,
\quad T_F(z) \psi^t(w) = {\p t(w)\over 2(z-w)} \, ,
\ee
\be\label{e3.18x}
\bar T_F(z) \bar \psi^3(w) =- {\bar J^3(w)\over 2(\bar z-\bar w)} , \quad \bar T_F(z) 
\bar \chi^3(w) = {\bar K^3(w)\over 2(\bar z-\bar w)} ,
\quad \bar T_F(z) \bar \psi^t(w) = {\bar \p t(w)\over 2(\bar z-\bar w)} \, .
\ee

\subsection{Partition function}

For computing the partition function of string theory on this target space, 
we make the identification
\be\label{e2.4xx}
(t,\vp, \phi_1,\phi_2) \equiv (t+2\pi\uptau_2, \vp+2\pi\uptau_1, \phi_1+2\pi \xi_1, \phi_2+2\pi \xi_2)\, ,
\ee
together with appropriate twists  in the $\psi^1$-$\psi^2$,
$\bar\psi^1$-$\bar\psi^2$, $\chi^1$-$\chi^2$,
$\bar\chi^1$-$\bar\chi^2$ planes so as to maintain world-sheet supersymmetry
and carry out the functional integral over the world-sheet fields with this twisted boundary
condition.
This can also be given an interpretation as a trace over the states of the
original untwisted world-sheet theory by regarding the system as an orbifold of the
original global $AdS_3$ by the transformation \refb{e2.4xx}. 
In this picture, while computing the world-sheet partition function 
we need to sum over winding modes where the fields
transform by \refb{e2.4xx} $(m',n')$ times as we move once along the $(a,b)$-cycle of the world-sheet torus.  Hence the total partition function is given by
\be\label{e2.6new}
\ln \ZB = \uptau_2 \times {1\over 2}\,  \int_\FF {d^2\tau\over \tau_2} \sum_{m',n'}\, Z^{m',n'}_{ws}\, ,
\ee
where $\FF$ is the region $-1/2\le\tau_1< 1/2$, $|\tau|>1$. The factor of $1/2$ is the usual factor associated
with the torus amplitude to account for the $z\to -z$ isometry of the world-sheet torus.
In the Euclidean path integral,
the $\uptau_2$ factor is the result of integration over the zero modes of the
scalar associated with the coordinate $t$.
By appropriate $SL(2,\ZZZ)$ transformation of the world-sheet torus, we can bring 
\refb{e2.6new} to
the form where the winding is $(0,n)$ where $n=\gcd(m',n')$ and at the 
same time add the contribution from the appropriate $SL(2,\ZZZ)$ images of the
original keyhole region $\FF$. This involves carrying out the
$\tau$ integral over the strip $\SSS$\cite{Polchinski:1985zf,2610.20696}
\be
\SSS: \qquad -{1\over 2}\le \tau_1<{1\over 2}, \qquad \tau_2>0\, .
\ee
The only exception is the $(m'=0,n'=0)$ 
sector contribution for which the integration runs over $\FF$. However for this sector the
contribution is proportional to $\uptau_2$ with no dependence on $\uptau_1$,
$\xi_1$ or $\xi_2$. Hence its effect can be removed by shifting the energies of all the 
states  by a constant and we shall ignore this contribution.
Then the  partition function $\ZB$ of string theory on this target space may be 
written as\cite{2610.20696}
\be\label{e2.6}
\ln \ZB = \uptau_2 \sum_{n=1}^\infty \int_\SSS {d^2\tau\over \tau_2} Z^{0,n}_{ws}\, ,
\ee
The factor of $1/2$ in \refb{e2.6new} has been cancelled against the factor of
2 coming from dropping the
sum over the negative values of $n$.
Let us introduce the complex parameters
\be
\uptau = \uptau_1+i\, \uptau_2, \qquad \xi = \xi_1 + \xi_2, \qquad \wt\xi = \xi_1 - \xi_2\, .
\ee
Then $Z_{ws}^{0,n}$ is given by
\be \label{e2.7}
Z_{ws}^{0,n}=Tr_{ws}\left[b_0c_0\bar b_0\bar c_0\,
e^{2\pi i n\uptau J^3_0 - 2\pi i n\bar\uptau \bar J^3_0 
+ 2\pi i n\xi K^3_0 - 2\pi i n\wt\xi \bar K^3_0+2\pi i \tau L_0-2\pi i \bar\tau\bar L_0}
\right]\, ,
\ee
where the trace is taken over all states in the original theory before orbifolding.
The subscript $0$ on the currents denotes their zero modes given by the integral along
the $a$-cycles of the torus. The precise normalization will be given in \refb{e3.13a}.

There are two points that are worth mentioning here.
\begin{enumerate}
\item $\ZB$ computed from \refb{e2.6} includes multi-string contribution. This is captured by having
to exponentiate the right hand side of \refb{e2.6} for computing $\ZB$.
\item In the Hamiltonian description \refb{e2.7},
the origin of the $\uptau_2$ factor in \refb{e2.6new} may be understood as follows.
If we regard the thermal $AdS_3$ as the quotient of the global $AdS_3$ by the orbifold group
$\mathbb{Z}$, then the 
full partition function is obtained by summing over all winding numbers and dividing
the result by the order of the orbifold group. Now the order of the orbifold group is
infinity but one can regulate it by taking the original target space as $AdS_3$ with the
time coordinate having a large period $\Lambda$ and then taking a quotient of this by
$\mathbb{Z}_{\Lambda/2\pi\uptau_2}$. The order of the orbifold group is now
$\Lambda/2\pi \uptau_2$. When we divide the partition function by this, we get a factor
of $2\pi\uptau_2/\Lambda$. The factor of $2\pi/\Lambda$ cancels against the factor of 
$\Lambda/2\pi$
that multiplies the density of states associated with the zero modes of the world-sheet field $t$,
and the full partition function is given by 
$\uptau_2 \, Z_{ws}$. 
\end{enumerate}

Since the result for $(0,n)$ winding number can be obtained from that of $(0,1)$ winding number
by the replacement $\uptau\to n\uptau$, $\xi\to n\xi$, 
$\bar\uptau\to n\bar\uptau$, $\wt\xi\to n\wt\xi$ in $Z_{ws}$,
we shall focus on the contribution from the
$(0,1)$ sector:
\be
f^{01}(\uptau,\bar\uptau,\xi,\wt\xi) = \uptau_2 \int_\SSS {d^2\tau\over \tau_2} Z^{0,1}_{ws}\, ,
\ee

Let us now examine how the problem described earlier shows up in the explicit computation.
We begin by writing some operator product expansions.
Defining
\be \label{e2.12}
\sigma^\pm = {1\over \sqrt k}\, (\chi^3 \pm \psi^3) % {1\over \sqrt k}\, (\psi^3\pm i \chi^3)\, ,
\ee
we get, from \refb{e2.5}
\be
\sigma^+(z) \sigma^-(w)\simeq  {1\over z-w}\, .
\ee

In the sector with spin-structure $\delta$, some relevant torus correlation functions are:
\be \label{e2.15}
\langle \sigma^+(z) \sigma^-(w)\rangle \propto \vt_\delta(z-w)\, {\vt_1'(0)\over \vt_1(z-w)}\, ,
\ee
\be \label{e2.15a}
\langle e^\phi(z) e^{-\phi}(w) \rangle \propto 
{1\over \vt_\delta(z-w)} \, {\vt_1(z-w)\over \vt_1'(0)}\, \, .
\ee
Here $\vt_\delta$ denote the Jacobi theta  functions.
Odd spin structure corresponds to $\delta=1$ and even spin structures correspond to
$\delta=2,3,4$. Similar results hold in the anti-holomorphic sector.

We can define the partition functions of the $\sigma^\pm$ and the $\beta$-$\gamma$ system
through the operator product expansion
\be
Z_{\sigma^\pm} = \lim_{z\to w} (z-w) \langle \sigma^+(z) \sigma^-(w)\rangle 
\propto \vt_\delta(0)
\ee
and
\be 
Z_{\beta\gamma} = \lim_{z\to w} (z-w)^{-1} \langle e^\phi(z) e^{-\phi}(w) \rangle 
\propto {1\over \vt_\delta(0)}\, .
\ee
This shows that for odd spin structure, corresponding to $\delta=1$, $Z_{\sigma^\pm}$
vanishes and $Z_{\beta\gamma}$ diverges, as expected. The same story repeats in the
anti-holomorphic sector. This can be traced to the existence of zero modes. In contrast the
modes $\psi^1\pm i\psi^2$ and $\chi^1\pm i\chi^2$ obey twisted boundary condition along
the $b$-cycle and do not lead to any ambiguity.

If instead of computing the partition function we compute the derivative of the partition function
with respect to $\uptau$, $\bar\uptau$, $\xi$ or $\wt\xi$, then we need to compute the one point
function of an appropriate closed string vertex operator on the torus. For the NS sector states
propagating in the loop, i.e. anti-periodic boundary condition along the $a$-cycle we can take
the vertex operator in the zero picture. However for R-sector states propagating in the loop,
corresponding to periodic boundary condition along the $a$-cycle, we need to choose the vertex
operator in the $-1$ picture and insert the average of the PCO along the $a$-cycle, -- the operator
$\XX_0$. This has two effects. First of all, we have an insertion of $e^{-\phi}$ at the location
of the vertex operator. Hence to get a non-zero result we need to pick the term in 
$\XX$ proportional to $e^\phi T_F$. Hence in the $\beta$-$\gamma$ sector we now have a
two point function $\langle e^\phi(z) e^{-\phi}(w) \rangle$ with $z\ne w$, and the contribution no
longer diverges. Furthermore, the vertex operator in the matter sector now supplies a factor
of $\sigma^+$ or $\sigma^-$, and we can get a non-zero result for the $\sigma^\pm$ correlator by
picking the term proportional to $\sigma^-$ or $\sigma^+$ from $T_F$ in the PCO insertion. As a
result we get a finite contribution to the correlator.

Even though the above discussion is valid for periodic boundary condition along the $a$-cycles
irrespective of the boundary condition along the $b$-cycle, if the $b$-cycle has anti-periodic
boundary condition then we can deform the PCO location $z$ and take it to  
the location $w$ of the vertex operator without encountering any spurious pole which is located
at the zero of $\vt_\delta(z-w)$ according to \refb{e2.15a}. Hence only for periodic boundary
condition along both cycles, i.e.\ for $\delta=1$, 
we need to insert the PCO at the location $z$ away from $w$.
Even though we are supposed to integrate $z$ along the $a$-cycle to construct the
zero mode $\XX_0$ of the PCO, the correlator will turn out
to be independent of $z$ and hence we can take $z$ to be at any point away from $w$. 
A similar remark holds for the anti-holomorphic sector.

Thus it remains to explicitly evaluate the correlation functions and evaluate the one point functions.
This will be done in the next section.

\sectiono{Computation of derivatives of the partition function} \label{s3}

In this section we shall compute the derivatives of the partition function with respect to various
chemical potentials.

\subsection{Vertex operators}

The strategy for constructing the vertex operator associated with the deformation of a chemical
potential (including temperature) is as follows. Each chemical potential is conjugate to a
charge that generates translation along an isometry direction. Let $\chi$ denote the isometry
direction
associated with the chemical potential $\mu$ where $\mu$ can be $\tau_1$, $\tau_2$,
$\xi_1$ or $\xi_2$. Then we have the 
identification\footnote{Note that the isometry not only acts on the angular directions of the
$AdS_3\times S^3$, but it also acts on the free fermions that form the superpartners of the 
$SL(2,R)\times SU(2)$ currents. This can be taken into account by bosonizing the free fermions
and including in the definition of $\chi$ the appropriate bosonic fields that arise from the
bosonization of the free fermions. The net effect of this is to include in the definition of the 
various currents appearing in \refb{e3.7} not only the contribution from the scalar fields associated
with the $SL(2,R)\times SU(2)$ group manifold, but also their fermionic partners.
}
\be \label{e3.3}
(t,\chi) \equiv (t+2\pi\uptau_2, \chi+2\pi \mu)\, .
\ee
A change in $\mu$ can be implemented by a transformation
\be \label{e3.4}
\delta\chi = {\delta\mu\over \uptau_2} \, t\, .
\ee
Since the world-sheet action $S_{ws}$ 
depends on $\chi$ only through the combination $\p_\alpha\chi$
where $\p_\alpha =\p / \p\zeta^\alpha$ denotes derivative with respect to the world-sheet
coordinate $\zeta^\alpha$, the change in $S_{ws}$ under the transformation \refb{e3.4} is
given by
\be 
 \int d^2\zeta\, {\delta S_{ws}\over \delta (\p_\alpha \chi)} \p_\alpha \left({\delta\mu\over \uptau_2} t\right)\sim
{\delta\mu\over \uptau_2} \,  \int d^2\zeta J^\alpha \p_\alpha t\, ,
 \ee
 where $J^\alpha$ is the conserved current that generates
 translation along $\chi$ and $\sim$ stands for up to overall normalization
 corresponding to a numerical coefficient.
This gives
\be\label{e3.7}
\delta S_{ws} \sim {1\over \uptau_2} \, \int d^2 \zeta \left[\delta \uptau \, J^3 \, \bar \p t + 
\eps_1\, \delta \bar \uptau  \, \p t \, \bar  J^3  - \eps_2 \, \delta \xi \, K^3 \, \bar \p t - \eps_3\,
\delta \wt\xi   \, \p t \, \bar  K^3 \right]
\ee
where the $\eps_i$'s are signs determined by the direction of translation generated by the
currents. The $\eps_i$'s and the overall normalization will be fixed later.

When the world-sheet is a torus, we shall take $\zeta=(x,y)$ and introduce complex coordinates
\be \label{e3.5a}
z = x+i\, y, 
\ee
with the identification
\be \label{e3.6a}
z\equiv z+1\equiv z+\tau, \qquad \tau=\tau_1+i\tau_2\, .
\ee
For later use we also record the derivatives with respect to $z$ and $\bar z$ in terms of
derivatives with respect to $x$ and $y$:
\be \label{e3.7a}
\p_z ={1\over 2} (\p_x - i\, \p_y), \qquad \p_{\bar z} 
={1\over 2} (\p_x + i\, \p_y)\, .
\ee

We need to find the $(-1,-1)$ picture unintegrated vertex operator for this deformation. 
Using \refb{e3.7}, \refb{e3.8} and \refb{e3.11} we can write down the $(-1,-1)$ picture vertex
operator associated with the deformation of $\uptau$, $\bar\uptau$, $\xi$ and $\wt\xi$:
\be
-{1\over 2\pi\uptau_2\, k} \, 
c \, \bar c\, e^{-\phi} \, e^{-\bar\phi} \, \left[
\delta \uptau \, \psi^3 \, \bar \psi^t + \eps_1
\delta \bar \uptau  \, \psi^t \, \bar  \psi^3  +\eps_2 \delta \xi \, \chi^3 \, \bar \psi^t + \eps_3
\delta \wt\xi   \, \psi^t \, \bar  \chi^3 
\right]\, ,
\ee
up to an overall normalization. This gives the following expression for the variation of the
partition function under this deformation:
\ben
\label{e3.17}
\delta f^{01}&=&  {\alpha\over  k} \,\int_\SSS d^2\tau\, 
 \bigg\langle b_0 c_0 \bar b_0 \bar c_0\, 
 \XX(z) \bar\XX(z')  \, e^{-\bar\phi}(w) \, e^{-\phi}(w) \, \nonumber \\ &&
 \bigg[
\delta \uptau \, \psi^3 \, \bar \psi^t(w) + \eps_1
\delta \bar \uptau  \, \psi^t \, \bar  \psi^3(w)   +\eps_2 \delta \xi \, \chi^3 \, \bar \psi^t(w)  + 
\eps_3
\delta \wt\xi   \, \psi^t \, \bar  \chi^3 (w) 
\bigg]\bigg\rangle\, ,
\een
where $\alpha$ is a normalization constant that will be fixed later.
Using $\phi$ charge conservation we see that the only parts of the picture changing operators
that will be relevant for our analysis are:
\be
\XX(z) = e^{\phi} T_F(z) +\cdots, \qquad \bar\XX(z') = e^{\bar\phi} \bar T_F(z')\, .
\ee

We shall end this subsection by discussing the normalization of the zero modes of the
currents that appear in \refb{e2.7}. 
\refb{enormint} implies that in the integration over $z$ we include a factor of $(2\pi i)^{-1}$ and in the
integration over $\bar z$ we include a factor of $(-2\pi i)^{-1}$.
For consistency, the same normalization convention \refb{enormint} must be used for
defining the zero modes of the currents. Since they are given by  integration along the
$a$-cycles, we get,
\ben\label{e3.13a}
&& J^3_0=-{1\over 2\pi i} \int_0^1 dz \, J^3(z), \qquad K^3_0=-{1\over 2\pi i}
\int_0^1 dz \, K^3(z), \nonumber \\ &&
\bar J^3_0={1\over 2\pi i} \int_0^1 d\bar z \, \bar 
J^3(z), \qquad \bar K^3_0={1\over 2\pi i} \int_0^1 d\bar z \, \bar K^3(z)\, .
\een
In fixing the signs we used the fact that in the $w=e^{-2\pi i (x+iy)}$ coordinate in which $y=-\infty$
is mapped  to
$w=0$ and $y=\infty$ is mapped
to $w=\infty$, the $a$-cycle that goes from $x=0$ to $x=1$ is mapped to a clockwise
contour in the $w$-plane.

\subsection{One point function}

We shall now evaluate the correlators that appear on the right hand side of \refb{e3.17}
for different choice of spin structures. We shall
analyze the contribution from different spin structure sectors separately.
For this we define
$\wh Z$ in different sectors via the relations\footnote{Note that in the 
odd spin structure there is an ambiguity in the sign 
of the correlation function since we have zero modes of $\psi^3$, $\chi^3$ and we have
to decide on the sign of $\langle \psi^3\chi^3\rangle$. A similar remark holds in the
anti-holomorphic sector. This is related to the GSO projection rules and the choice
of sign differs between type IIA and type IIB theories.}
\be \label{e3.23}
\wh Z_{ee} = \langle b_0c_0\bar b_0\bar c_0\rangle_{ee}\, ,
\ee
\be  \label{e3.24}
\wh Z_{oo}=  \bigg\langle 
b_0 c_0 \bar b_0 \bar c_0\, e^{\phi(z)} 
e^{-\phi(w)} \sigma^+(z) \sigma^-(w)
e^{\bar\phi(z')} 
e^{-\bar\phi(w)} \bar\sigma^+(z') \bar\sigma^-(w)
\bigg\rangle_{oo}\, ,
\ee
\be \label{e3.25}
\wh Z_{eo}=  \bigg\langle b_0 c_0 \bar b_0 \bar c_0\, e^{\phi(z)} 
e^{-\phi(w)} \sigma^+(z) \sigma^-(w)
\bigg\rangle_{eo}\, ,
\ee
\be \label{e3.26}
\wh Z_{oe}=   \bigg\langle b_0 c_0 \bar b_0 \bar c_0\, e^{\bar\phi(z')} 
e^{-\bar\phi(w)} \bar\sigma^+(z') \bar\sigma^-(w)
\bigg\rangle_{oe}\, .
\ee
Due to \refb{e2.15}, \refb{e2.15a} these expressions are independent of $z,z',w$ and their
anti-holomorphic counterparts.

\subsubsection{Even - even spin structure}

In the even spin structure sector we can take $z\to w$, $\bar z'\to \bar w$ limit and 
get
\be
\delta f^{01}_{ee} = {\alpha\over k} \, 
{1\over 4}\,  \int_\SSS d^2\tau\, 
 \bigg\langle b_0 c_0 \bar b_0 \bar c_0\, 
 \bigg[
-\delta \uptau \, J^3 \, \bar \p t - \eps_1
\delta \bar \uptau  \, \p t \, \bar  J^3  +\eps_2 \delta \xi \, K^3 \, \bar \p t +\eps_3
\delta \wt\xi   \, \p t \, \bar  K^3
\bigg]\bigg\rangle_{ee} \, ,
\ee
with the understanding that the $z$ and $w$ and $z'$ and $w$ are close but not equal.
We have shown in appendix \ref{sa} that 
inside the correlation function we can make the replacement:
\ben\label{e3.29}
&& J^3\, \bar\p t = {i\pi \uptau_2\over \tau_2} \, J^3-  {\pi\over 2\tau_2}, \qquad 
\p t  \,
\bar J^3= -{i\pi \uptau_2\over \tau_2} \,
\bar J^3 -  {\pi\over 2\tau_2}\, , \nonumber \\
&& K^3 \, \bar\p t = {i\pi \uptau_2\over\tau_2} \, K^3, \qquad  \p t \, \bar K^3
= -{i\, \pi \, \uptau_2\over \tau_2} \,
\bar K^3 \, .
\een
Using \refb{e3.13a}, \refb{e3.29}
we now get
\ben \label{e3.27aa}
\delta  f^{01}_{ee}  &=& {\alpha\over 4 \, k} \,  \int_\SSS {d^2\tau\over \tau_2}
\,  \bigg\langle b_0 c_0 \bar b_0 \bar c_0\, \pi \, i\, \uptau_2  \, 2\pi i\, 
 \bigg[
\delta \uptau \, J^3_0 + \eps_1
\delta \bar \uptau  \, \bar  J^3_0  -\eps_2 \delta \xi \, K^3_0 \,  - \eps_3
\delta \wt\xi   \,\bar  K^3_0
\bigg]\bigg\rangle_{ee} \nonumber \\
&+ &{\alpha\, \pi \over 4 \, k
} \, \left({1\over 2}\right)\,  \int_\SSS {d^2\tau\over \tau_2}
\, \bigg\langle b_0 c_0 \bar b_0 \bar c_0\, 
 \bigg[
\delta \uptau \ + \eps_1
\delta \bar \uptau    
\bigg]\bigg\rangle_{ee}  \, .
\een
Using \refb{e2.7} and \refb{e3.23}, \refb{e3.27aa} can be written as,
\ben \label{e3.20yx}
\delta  f^{01}_{ee}  &=& -{\alpha\over 4\, k} \, 2 \pi \uptau_2 \, 
  {i\over 2}\, \bigg[-
\delta \uptau \, {\p\over \p\uptau} +\eps_1
\delta \bar \uptau  \, {\p\over \p\bar\uptau} +\eps_2 \delta \xi \, {\p\over \p\xi} \,  - \eps_3
\delta \wt\xi   \, {\p\over \p\wt\xi}
\bigg]
 \int_\SSS {d^2\tau\over \tau_2}
\,  \wh Z_{ee} \nonumber \\
&+& {\alpha\, \pi \over 4 \, k
} \, {1\over 2}\bigg[
\delta \uptau \ + \eps_1
\delta \bar \uptau    
\bigg]\int_\SSS {d^2\tau\over \tau_2}
\, \wh Z_{ee}  \, .
\een
Since there are no subtleties in the even spin structure sector, we can 
equate this to the expected result:
\be \label{e3.21yx}
\delta  f^{01}_{ee}  =
\bigg[
\delta \uptau \, {\p\over \p\uptau} + 
\delta \bar \uptau  \, {\p\over \p\bar\uptau} + \delta \xi \, {\p\over \p\xi} \,  + 
\delta \wt\xi   \, {\p\over \p\wt\xi}
\bigg] \bigg[ \uptau_2
 \int_\SSS {d^2\tau\over \tau_2}
\,  \wh Z_{ee} \bigg]\, ,
\ee
to get,
\be
\alpha={4\, k\over i\pi}, \qquad  \eps_1=\eps_2=-1, \qquad \eps_3=1\, .
\ee

This can be used in \refb{e3.17} to write the expression for the first order variation of the partition
function in general spin structure:
\ben
\delta \, f^{01} &=&  {4\over i\pi} \,\int_\SSS d^2\tau\, 
 \bigg\langle b_0 c_0 \bar b_0 \bar c_0\, 
 \XX(z) \bar\XX(z')  \, e^{-\bar\phi}(w) \, e^{-\phi}(w) \, \nonumber \\ &&
 \bigg[
\delta \uptau \, \psi^3 \, \bar \psi^t -
\delta \bar \uptau  \, \psi^t \, \bar  \psi^3  -\delta \xi \, \chi^3 \, \bar \psi^t +
\delta \wt\xi   \, \psi^t \, \bar  \chi^3 
\bigg]\bigg\rangle\, .
\een
We note in passing that in order that \refb{e3.20yx} and \refb{e3.21yx} agree, it is important that
the terms in the second line of \refb{e3.20yx} come with the correct coefficients. This in 
turn can be
traced to the $-\pi/2\tau_2$ terms in \refb{e3.29}, providing independent evidence that these
coefficients are correct. 

We shall now analyze the contribution from the odd-odd, odd-even and even-odd spin structure
sectors. 
\subsubsection{Odd - odd spin structure}

We have
\ben
\delta  f^{01}_{oo} &=&  {4\over \pi i  } \,\int_\SSS d^2\tau\, 
 \bigg\langle b_0 c_0 \bar b_0 \bar c_0\, 
 \XX(z) \bar\XX(z')  \, e^{-\bar\phi}(w) \, e^{-\phi}(w) \, \nonumber \\ &&
 \bigg[
\delta \uptau \, \psi^3 \, \bar \psi^t (w)-
\delta \bar \uptau  \, \psi^t \, \bar  \psi^3(w)  -\delta \xi \, \chi^3 \, \bar \psi^t(w) +
\delta \wt\xi   \, \psi^t \, \bar  \chi^3 (w)
\bigg]\bigg\rangle_{oo}\, .
\een
This can be written as
\ben
\delta  f^{01}_{oo} &=&  {4\over \pi i } \,\int_\SSS d^2\tau\, 
 \bigg\langle b_0 c_0 \bar b_0 \bar c_0\, 
e^{\phi}(z) \, e^{-\phi}(w) \, e^{\bar\phi}(z') e^{-\bar\phi}(w) \,  \nonumber \\ &&
 \bigg[
\delta \uptau \, T_F(z) \, \psi^3(w) \,  \bar T_F(z') \, \bar \psi^t (w)-
\delta \bar \uptau  \, T_F(z) \, \psi^t(w) \, \bar T_F(z') \,\bar  \psi^3 (w) \nonumber \\
&& -\delta \xi \, T_F(z) \, 
\chi^3 (w)\, \bar T_F(z') \, \bar \psi^t (w) +
\delta \wt\xi   \, T_F(z) \, \psi^t (w)\,\bar T_F(z') \, \bar  \chi^3 (w)
\bigg]\bigg\rangle_{oo}\, .
\een
We need to insert in the correlator a factor of $\psi^3 \chi^3\bar\psi^3\bar\chi^3$ to get a
non-zero answer. Hence using \refb{e3.11a} and \refb{e3.11} we see that the only non-vanishing 
contribution is:
\ben\label{e3.33}
\delta  f^{01}_{oo} &=&  {4\over \pi i \, k} \,{1\over k^2}\times 
\left(-1\right) \,
\int_\SSS d^2\tau\, 
 \bigg\langle b_0 c_0 \bar b_0 \bar c_0\, 
e^{\phi}(z) \, e^{-\phi}(w) \, e^{\bar\phi}(z') e^{-\bar\phi}(w) \,  \nonumber \\ &&
 \bigg[
\delta \uptau \, K^3\chi^3(z) \, \psi^3(w) \,  \bar K^3 \bar\chi^3(z') \, \bar \psi^3 (w)-
\delta \bar \uptau  \, K^3\chi^3(z) \, \psi^3(w) \, \bar K^3 \bar\chi^3(z') \,\bar  \psi^3 (w) \nonumber \\
&& -\delta \xi \, J^3\psi^3(z) \, 
\chi^3 (w)\, \bar K^3\bar\chi^3(z') \, \bar \psi^3 (w) +
\delta \wt\xi   \, K^3\chi^3(z) \, \psi^3 (w)\,\bar J^3\bar\psi^3(z') \, \bar  \chi^3 (w)
\bigg]\bigg\rangle_{oo}\, . \nonumber \\
\een
Using \refb{e2.12} we get
\be \label{e3.38y}
\psi^3(z) \chi^3 (w)  =  {k\over 4} % {k\over 4 i} 
(\sigma^+(z) -\sigma^-(z)) (\sigma^+(w) + \sigma^-(w))
=  {k\over 4} % -{k\over 4i} 
(\sigma^+(z) \sigma^-(w) + \sigma^+(w) \sigma^-(z) +\cdots)\, ,
\ee
where $\cdots$ involve products of two $\sigma^+$ or two $\sigma^-$ and hence do not
contribute to the correlation function. Similarly we have,
\be \label{e3.38yxxz}
\bar\psi^3(z') \bar\chi^3 (w)  =
{k\over 4} % -{k\over 4i} 
(\bar\sigma^+(z') \bar\sigma^-(w) + \bar\sigma^+(w) \bar\sigma^-(z') +\cdots)\, ,
\ee
Now it follows from \refb{e2.15}, \refb{e2.15a} that in the odd spin structure sector,
$\sigma^+(z) \sigma^-(w)$ and $\sigma^+(w) \sigma^-(z)$ have the same correlation functions.
Using these we can express \refb{e3.33} as
\ben
\delta  f^{01}_{oo} &=&  {4\over \pi i \, k} \,{1\over k^2}\,
{k^2\over 16} \times 4\, \left(-1\right)\, 
\int_\SSS d^2\tau\, 
 \bigg\langle b_0 c_0 \bar b_0 \bar c_0\, 
 e^{\phi}(z) \, e^{-\phi}(w) \, e^{\bar\phi}(z') e^{-\bar\phi}(w) 
  \nonumber \\ &&
 \sigma^+(z) \sigma^-(w)\bar\sigma^+(z') \bar\sigma^-(w) 
   \nonumber \\ && \bigg[
\delta \uptau \, K^3(z) \,  \bar K^3 (z') 
 -
\delta \bar \uptau  \, K^3(z) \, \bar K^3(z')  +\delta \xi \, J^3(z) \, 
\bar K^3(z') -
\delta \wt\xi   \, K^3(z) \, \bar J^3(z') 
\bigg]\bigg\rangle_{oo}\, . \nonumber \\
\een
Using \refb{e2.7} and \refb{e3.24}, this can be rewritten as
\be \label{e3.41x}
\delta  f^{01}_{oo} = - {1\over \pi \, i\, k} 
 \int_\SSS d^2\tau\, 
  \bigg[
\delta \uptau \, {\p^2\wh Z_{oo}\over \p\xi\p\wt\xi} 
-
\delta \bar \uptau  \, {\p^2\wh Z_{oo}\over \p\xi\p\wt\xi}  +\delta \xi \, 
{\p^2\wh Z_{oo}\over \p\uptau\p\wt\xi} 
 -
\delta \wt\xi   \, 
{\p^2\wh Z_{oo}\over \p\bar\uptau\p\xi} 
\bigg]\, .
\ee

\subsubsection{Even - odd spin structure}

We have
\ben
\delta f^{01}_{eo} &=&  {4\over i\, \pi } \,\int_\SSS d^2\tau\, 
 \bigg\langle b_0 c_0 \bar b_0 \bar c_0\, 
 \XX(z) \bar\XX(z')  \, e^{-\bar\phi}(w) \, e^{-\phi}(w) \, \nonumber \\ &&
 \bigg[
\delta \uptau \, \psi^3 \, \bar \psi^t (w)- 
\delta \bar \uptau  \, \psi^t \, \bar  \psi^3(w)  -\delta \xi \, \chi^3 \, \bar \psi^t(w) +
\delta \wt\xi   \, \psi^t \, \bar  \chi^3 (w)
\bigg]\bigg\rangle_{eo}\, .
\een
Since the spin-structure in the anti-holomorphic (left-chiral) sector is even, we can take
$\bar z'\to \bar w$ limit using \refb{e3.18x} to get
\ben
\delta f^{01}_{eo} &=&  -{4\over i\pi} \,
{1\over 2} \int_\SSS d^2\tau\, 
 \bigg\langle b_0 c_0 \bar b_0 \bar c_0\, 
 e^\phi T_F(z) \, e^{-\phi}(w) \, \nonumber \\ &&
 \bigg[
\delta \uptau \, \psi^3 \, \bar \p t (w) +
\delta \bar \uptau  \, \psi^t \, \bar  J^3(w)  -\delta \xi \, \chi^3 \, \bar \p t(w) +
\delta \wt\xi   \, \psi^t \, \bar  K^3 (w)
\bigg]\bigg\rangle_{eo}\, ,
\een
where the extra minus sign comes from having to move $\bar T_F$ through the other
operators.
It is again understood that the PCOs are close to the vertex but not on top of the vertex so that
we do not pick up any contribution from the delta function contact terms.
Using \refb{e3.11a}, the non-vanishing part of the correlation function may be expressed as,
\ben
\delta f^{01}_{eo} &=&  {4\over i\pi \, k} \, 
{1\over 2} \int_\SSS d^2\tau\, 
 \bigg\langle b_0 c_0 \bar b_0 \bar c_0\, 
 e^\phi(z) \, e^{-\phi}(w) \, \nonumber \\ &&
 \bigg[
\delta \uptau \, \chi^3 K^3 (z)\psi^3(w) \, \bar \p t (w) +
\delta \bar \uptau  \, \chi^3 K^3 (z) \psi^t(w) \, \bar  J^3(w)  \nonumber \\ &&
-\delta \xi \, \psi^3 J^3 (z)\chi^3(w) \, \bar \p t(w) +
\delta \wt\xi   \, \chi^3 K^3 (z) \psi^t(w) \, \bar  K^3 (w)
\bigg]\bigg\rangle_{eo}\, .
\een
Using \refb{e3.11}, \refb{e3.38y} and the fact that in the odd spin structure sector,
$\sigma^+(z) \sigma^-(w)$ and $\sigma^+(w) \sigma^-(z)$ 
have the same correlation functions, we get
\ben
\delta f^{01}_{eo} &=&  {4\over i\, \pi \, k} \, 
{1\over 2} {k\over 2} \int_\SSS d^2\tau\, 
 \bigg\langle b_0 c_0 \bar b_0 \bar c_0\,  e^\phi(z) \, e^{-\phi}(w) \, \sigma^+(z) \sigma^-(w)
   \nonumber \\ &&\bigg[-
\delta \uptau \, K^3 (z) \, \bar \p t (w)+ {1\over k}\, 
\delta \bar \uptau  \,  K^3 (z)  \, \bar  J^3(w)
-\delta \xi \, J^3 (z) \bar \p t(w) +{1\over k}
\delta \wt\xi    K^3 (z) \, \bar  K^3 (w)
\bigg]\bigg\rangle_{eo}\, . \nonumber \\
\een
Following the same procedure used in the analysis of $\wh Z_{ee}$ and $\wh Z_{oo}$,
this can be written as,
\be
\delta f^{01}_{eo} =  {1\over i \pi}\int_\SSS d^2\tau\, 
  \bigg[
\delta \uptau \, {i\pi\uptau_2\over \tau_2} {\p \wh Z_{eo}\over \p \xi}+ {1\over k} 
\delta \bar \uptau  \,   {\p^2 \wh Z_{eo}\over \p\xi \p \bar\uptau}
+\delta \xi \, \, {i\pi\uptau_2\over \tau_2} {\p \wh Z_{eo}\over \p \uptau} 
+\delta\xi \, {\pi\over 2\tau_2}  \wh Z_{eo}
+{1\over k}
\delta \wt\xi    \,  {\p^2 \wh Z_{eo}\over \p\xi \p \wt\xi}
\bigg]\, .
\ee
This can be rewritten as
\be \label{e3.47x}
\delta f^{01}_{eo} =  {1\over i\,  \pi}\int_\SSS d^2\tau\, 
  \bigg[
\delta \uptau \, {i\pi\uptau_2\over \tau_2} {\p \wh Z_{eo}\over \p \xi}+{1\over k} 
\delta \bar \uptau  \,   {\p^2 \wh Z_{eo}\over \p\xi \p \bar\uptau}
+\delta \xi \, \, {i\pi\over \tau_2} {\p \over \p \uptau} (\uptau_2 \wh Z_{eo})
+{1\over k}
\delta \wt\xi    \,  {\p^2 \wh Z_{eo}\over \p\xi \p \wt\xi}
\bigg]\, . 
\ee

\subsubsection{Odd - even spin structure}

We have
\ben
\delta f^{01}_{oe} &=&  {4\over i\pi } \,\int_\SSS d^2\tau\, 
 \bigg\langle b_0 c_0 \bar b_0 \bar c_0\, 
 \XX(z) \bar\XX(z')  \, e^{-\bar\phi}(w) \, e^{-\phi}(w) \, \nonumber \\ &&
 \bigg[
\delta \uptau \, \psi^3 \, \bar \psi^t (w)-
\delta \bar \uptau  \, \psi^t \, \bar  \psi^3(w)  -\delta \xi \, \chi^3 \, \bar \psi^t(w) +
\delta \wt\xi   \, \psi^t \, \bar  \chi^3 (w)
\bigg]\bigg\rangle_{oe}\, .
\een
Since the spin-structure in the holomorphic (right-chiral) sector is even, we can take
$z\to w$ limit and use \refb{e3.17x} to get
\ben
\delta f^{01}_{oe} &=&  -{4\over i\pi} \, {1\over 2}\, 
\int_\SSS d^2\tau\, 
 \bigg\langle b_0 c_0 \bar b_0 \bar c_0\, 
 e^{\bar\phi} \bar T_F(z') \, e^{-\bar\phi}(w) \, \nonumber \\ &&
 \bigg[-
\delta \uptau \, J^3 \bar\psi^t (w)-
\delta \bar \uptau  \, \p t \bar \psi^3(w)  -
\delta \xi   \, K^3 \bar \psi^t  (w) + \delta \wt\xi \,  \p t(w)\, \bar \chi^3 (w)\,
\bigg]\bigg\rangle_{oe}\, .
\een
Using \refb{e3.11a}, the non-vanishing part of the correlation function may be expressed as,
\ben
\delta f^{01}_{oe} &=&  {4\over i\pi k} \, {1\over 2}\, 
\int_\SSS d^2\tau\, 
 \bigg\langle b_0 c_0 \bar b_0 \bar c_0\, 
 e^{\bar\phi}(z') \, e^{-\bar\phi}(w) \, \nonumber \\ &&
 \bigg[-
\delta \uptau  \, J^3(w) \bar \chi^3 \bar K^3(z')  \bar \psi^t(w)  -
\delta  \bar \uptau \, \, \p t (w)  \bar \chi^3 \bar K^3 (z')\bar \psi^3(w)  \nonumber \\ &&
-  \delta \xi   \,    K^3 (w)\, \bar\chi^3 \bar K^3 (z') \bar \psi^t(w)
+ \delta \wt\xi \, \p t(w) \, \bar \psi^3 \bar J^3 (z') \bar \chi^3(w) 
\bigg]\bigg\rangle_{oe}\, .
\een
Using \refb{e3.11}, \refb{e3.38yxxz} and the fact that in the odd spin structure sector,
$\bar\sigma^+(z) \bar\sigma^-(w)$ and $\bar\sigma^+(w) \bar\sigma^-(z)$ 
have the same correlation functions, we get
\ben
\delta f^{01}_{oe} &=&  {4\over i\pi\, k} \, 
{1\over 2} {k\over 2}  \int_\SSS d^2\tau\, 
 \bigg\langle b_0 c_0 \bar b_0 \bar c_0\, e^{\bar\phi}(z') \, e^{-\bar\phi}(w) \,
 \bar\sigma^+(z')\bar\sigma^-(w)\, \nonumber \\ &&
 \bigg[-
\delta  \uptau  \, {1\over k} \, J^3(w) \, 
\bar K^3 (z')  + \delta \bar \uptau  \,  \p t (w)\, \bar K^3 (z') 
 \nonumber \\ &&
- \delta  \xi   {1\over k}\,  K^3 (w)\,  \bar K^3 (z') + \delta \wt\xi \,  \p t(w)\, \bar J^3 (z')
\bigg]\bigg\rangle_{oe}\, .
\een
Following the same procedure used in the analysis of $\wh Z_{ee}$ and $\wh Z_{oo}$,
this can be written as,
\be \label{e3.52x}
\delta f^{01}_{oe} =  {1\over i\, \pi}\int_\SSS d^2\tau\, 
  \bigg[-\delta \uptau  \,  {1\over k}  {\p^2 \wh Z_{oe}\over \p\wt\xi \p \uptau} +
\delta \bar\uptau \, {i\pi\uptau_2\over \tau_2} {\p \wh Z_{oe}\over \p \wt\xi} 
- {1\over k}\, 
\delta \xi    \,   {\p^2 \wh Z_{oe}\over \p\wt\xi \p \xi}
+ \delta \wt\xi \, {i\pi\over \tau_2} {\p \over \p \bar\uptau}
(\uptau_2  \wh Z_{oe})\bigg]\, .
\ee

\sectiono{Comparison with the results of EGMR} \label{s4}

\newcommand{\Zh}{\widehat Z}

\newtheorem{prop}{Proposition}

\newcommand{\sg}{\epsilon_L}

\newtheorem{lemma}{Lemma}

In this section we shall compare our results with that of \cite{2610.20696}.

\subsection{Results for $\wh Z$}

We shall begin by extracting the results for $\widehat Z$, which were called naive partition function in
\cite{2610.20696}. We define
\be
\bm q=e^{2\pi i\bt}, \qquad \bm y=e^{2\pi i\bxi}, \qquad \widetilde{\bm y}=e^{-2\pi i\xit}, \qquad
q = e^{2\pi i\tau}\, ,
\ee
\ben \label{e4.2}
&& P(\bt,\tau)\equiv\sqrt{\frac{k}{\mathrm{Im}\,\tau}}\;e^{-\frac{k\pi(\mathrm{Im}\,\bt)^2}{\mathrm{Im}\,\tau}},\qquad
L_{\ell,m}\equiv \ell+km+\tfrac12, \nonumber \\ 
&& \Theta_{k,\ell}(\bxi\,|\,\tau)=\sum_{m\in\mathbb Z}q^{\frac1k L_{\ell,m}^2}
\bm y^{{L_{\ell,m}}},\nonumber\\
&&
\Theta^-_\ell(\bxi|\tau)\equiv\Theta_{k,\ell}(\bxi|\tau)-\Theta_{k,\ell}(-\bxi|\tau)
\ =\sum_m q^{\frac1kL_{\ell,m}^2}\big(\bm y^{{L_{\ell,m}}}-\bm y^{-{L_{\ell,m}}}\big),\nonumber \\ 
&&\mathcal C_\ell(\bxi|\tau)=\frac{\Theta^-_\ell(\bxi|\tau)}{\vartheta_1(\bxi|\tau)}\, .
% \qquad  \widetilde\Theta^-_\ell\equiv\Theta_{k,\ell}(-\xit|-\bar\tau)-\Theta_{k,\ell}(\xit|-\bar\tau).
\een
The sum over $\ell$ runs over $\ell\in\{0,\tfrac12,\dots,\tfrac k2-1\}$. We also define,
\be
\mathcal Z^{\rm K3}_{1,1}(\tau)=\chi(\mathrm{K3})=24, \qquad 
\mathcal Z^{\rm K3}_{a,1}(\tau)=\varphi^{\rm K3}_a(0|\tau), \qquad a=2,3,4\, , 
\ee
where
\be
\varphi^{\rm K3}_2(0|\tau)=-8\Big(\tfrac{\vartheta_4^2}{\vartheta_3^2}+\tfrac{\vartheta_3^2}{\vartheta_4^2}\Big),\quad
\varphi^{\rm K3}_3(0|\tau)=8\Big(\tfrac{\vartheta_4^2}{\vartheta_2^2}-\tfrac{\vartheta_2^2}{\vartheta_4^2}\Big),\quad
\varphi^{\rm K3}_4(0|\tau)=8\Big(\tfrac{\vartheta_3^2}{\vartheta_2^2}+\tfrac{\vartheta_2^2}{\vartheta_3^2}\Big),
\ee
with $\vartheta_\mu\equiv\vartheta_\mu(0|\tau)$. In the anti-holomorphic sector we have
\be
\mathcal Z^{\rm K3}_{1,\at}(\tau)=\varphi^{\rm K3}_{\at}(0|-\bar\tau)\, .
\ee
Then we have
\be
\widehat Z_{oo}
=6\,\alpha\alt\;P(\bt,\tau)\sum_\ell\Theta^-_\ell(\bxi|\tau)\, \Theta^-_\ell(-\wt\xi|-\bar\tau)\, ,
\ee
where $\alpha$ and $\wt\alpha$ take values $\pm 1$ depending on whether we are
considering type IIA or type IIB string theory. We also have
\be
\widehat Z_{eo}=-\frac{\alpha}{4}\,P(\bt,\tau)\sum_\ell\Theta^-_\ell(\bxi|\tau)\, {\mathcal C}_\ell
(-\wt\xi|-\bar\tau)\;
\frac{\displaystyle\sum_{\at=2,3,4}\alt_{\at}\,\vartheta_{\at}(-\bbt|-\bar\tau)\,\vartheta_{\at}(-\xit|-\bar\tau)\,
\varphi^{\rm K3}_{\at}(0|-\bar\tau)}{\vartheta_1(-\bbt|-\bar\tau)}\, ,
\ee
\be
\wh Z_{oe} = -\frac{\alt}{4}\,P(\bt,\tau)\sum_\ell\mathcal C_\ell(\bxi|\tau)\,\Theta^-_\ell (-\wt\xi|-\bar\tau)\;
\frac{\displaystyle\sum_{a=2,3,4}\alpha_a\,\vartheta_a(\bt|\tau)\,\vartheta_a(\bxi|\tau)\,\varphi^{\rm K3}_{a}(0|\tau)}{\vartheta_1(\bt|\tau)}\, ,
\ee
where
\be
\alpha_2=-1, \quad \alpha_3=+1, \quad \alpha_4=-1, \quad \wt\alpha_2=-1, \quad \wt\alpha_3=+1, \quad 
\wt\alpha_4=-1\, .
\ee

\subsection{Proposal for $f^{01}$}

We first define:
\ben
\Psi^-_\ell(\xi|\tau) &=& \Theta^-_\ell(\xi|\tau) = \sum_{m\in \mathbb Z} q^{{1\over k}L_{\ell,m}^2} (\by^{L_{\ell,m}} - \by^{-{L_{\ell,m}}}),
\nonumber \\
 \Psi^+_\ell(\xi|\tau) &=& \sum_{m\in \mathbb Z}
\mathrm{sgn}({L_{\ell,m}})\,q^{\frac1kL_{\ell,m}^2}\big(\bm y^{|{L_{\ell,m}}|}+\bm y^{-|{L_{\ell,m}}|}\big)\, ,
\een
where ${L_{\ell,m}}$ has been defined in \refb{e4.2}. We now define
$\wh Z^{EGMR}$ from $\wh Z$ by the following replacement rules:
\begin{enumerate}
\item in every \emph{holomorphic odd} factor we replace $\Theta^-_\ell(\xi|\tau)$ by $\Psi^+_\ell(
\xi|\tau)$;
\item in every \emph{antiholomorphic odd} factor we replace
$\Theta^-_\ell(-\wt\bxi|-\bar\tau)$ by $\Psi^+_\ell(-\xit|-\bar\tau)$;
\item the even--even sector is unchanged.
\end{enumerate}
Then the proposed expression for $f^{01}$ is:
\be
f^{01}_{xy}=\bt_2\, \int_\SSS\frac{d^2\tau}{\tau_2}\,\widehat Z^{\rm EGMR}_{xy}\, ,
\ee
where $x,y$ can take values $e$ or $o$. In particular, we have\footnote{These integrals
have divergences associated with the long string sector, but this has already been discussed
in \cite{2610.20696} and we do not have anything new to add.}
\be
f^{01}_{oo}=6\, \alpha\, \wt\alpha\, 
\bt_2\, \!\int_\SSS\!\frac{d^2\tau}{\mathrm{Im}\,\tau}\,P(\bt,\tau)\sum_\ell\Psi^+_\ell(\bxi|\tau)\,
\Psi^+_\ell(-\wt\xi|-\bar\tau),\label{eq:intoo}
\ee
\ben
f^{01}_{eo} &=&-\frac14\, \alpha\, \bt_2\, \!\int_\SSS\!\frac{d^2\tau}{\mathrm{Im}\,\tau}\,P(\bt,\tau)\sum_\ell\Psi^+_\ell(\bxi|\tau)\,
{\mathcal C}_\ell(-\wt\xi|-\bar\tau)
\nonumber \\
&& \frac{\sum_{\at=2,3,4}\alt_{\at}\,\vartheta_{\at}(-\bbt|-\bar\tau)\vartheta_{\at}(-\xit|-\bar\tau)\,
\varphi^{\rm K3}_{\at}(0|-\bar\tau)}{\vartheta_1(-\bbt|-\bar\tau)},\label{eq:inteo}
\een
\be
f^{01}_{oe}=-\frac14\, \wt\alpha\, \bt_2\, \int_\SSS\!\frac{d^2\tau}{\mathrm{Im}\,\tau}\,P(\bt,\tau)\sum_\ell\mathcal C_\ell(\bxi|\tau)\,
\Psi^+_\ell(-\wt\xi|-\bar\tau)\;\frac{\sum_{a=2,3,4}\alpha_a\,\vartheta_a(\bt|\tau)\vartheta_a(\bxi|\tau)\,\varphi^{\rm K3}_a(0|\tau)}{\vartheta_1(\bt|\tau)}\, .
\ee
Our goal will be to check that the $f^{01}$'s defined above satisfy \refb{e3.41x}, \refb{e3.47x}, \refb{e3.52x}.

\subsection{Comparison}

We write
\be
\Theta^-_\ell(\bxi|\tau)=\sum_m q^{L_{\ell,m}^2/k}\,\theta_{L_{\ell,m}}(\bxi),\qquad \Psi^+_\ell(\bxi|\tau)=\sum_m q^{{L_{\ell,m}^2/k}}\,\psi_{L_{\ell,m}}(\bxi),
\ee
where
\be
\theta_{L}\equiv\by^{L}-\by^{-{L}},\qquad \psi_{L}\equiv
\mathrm{sgn}({L})
\,(\by^{|{L}|}+\by^{-|{L}|})=\mathrm{sgn}({L})\,(\by^{L}+\by^{-{L}}).
\ee
We now collect some useful relations for proving the desired results.
\begin{enumerate}
\item 
A very useful relation is:
\begin{equation}\label{swap}
\partial_{\bxi}\theta_L=2\pi i|L|\,\psi_L,\quad \partial_{\bxi}\psi_L=2\pi i|L|\,\theta_L\, .
\end{equation}
From this it follows that
\be
f^{01}_{eo}=\frac{1}{i\pi k}\int_\SSS d^2\tau\,\partial_{\bxi}\Zh_{eo},\qquad f^{01}_{oe}=-\frac{1}{i\pi k}\int_\SSS d^2\tau\,\partial_{\xit}\Zh_{oe}.
\ee
\item 
Another useful observation is that since the integration over $\tau$ is carried out over the strip, the
$\tau_1$ integral gives level matching and only the $\tau_2$ integral remains. 
\item We have the relations
\begin{align}
&\int_\SSS  d^2\tau\,P(\bt,\tau)\,(q\bar q)^{L^2/k}=\frac{k}{2|L|}\,E_L, \label{J0}\\
&\int_\SSS d^2\tau\,\frac{\bt_2}{\tau_2}\,P(\bt,\tau)\,(q\bar q)^{L^2/k}=E_L, \label{J1}\\
&\int_\SSS d^2\tau\,\partial_{\bt}P(\bt,\tau)\,(q\bar q)^{L^2/k}=ik\pi\,E_L,\qquad \int_\SSS d^2\tau\,\partial_{\bbt}P(\bt,\tau)\,(q\bar q)^{L^2/k}=-ik\pi\,E_L, \label{J2}\\
&\int_\SSS \frac{d^2\tau}{\tau_2}\,\partial_{\bt}\big(\bt_2P(\bt,\tau)\big)(q\bar q)^{L^2/k}=2\pi i|L|\,E_L,
\nonumber \\ &
\int_\SSS \frac{d^2\tau}{\tau_2}\,\partial_{\bbt}\big(\bt_2P(\bt,\tau)\big)(q\bar q)^{L^2/k}=-2\pi i|L|\,E_L\, , \label{J3}
\end{align}
where,
\be
E_L\equiv e^{-4\pi\bt_2|L|}=(\bm q\bar{\bm q})^{|L|}\, .
\ee
\end{enumerate}
Using these relations it is straightforward to verify \refb{e3.41x}, \refb{e3.47x}, \refb{e3.52x}.

\bigskip

\noindent{\bf Acknowledgement:} 
I wish to thank Lorenz Eberhardt, Matthias Gaberdiel, Sameer Murthy,  Mukund Rangamani and Bogdam
Stefanski for 
useful discussions.
This work was supported by the ICTS-Infosys Madhava 
Chair Professorship
and the Department of Atomic Energy, Government of India, under project no. RTI4019.

\paragraph{Statement on AI use.} The analysis presented in this paper used
assistance from AI tool  Claude.
The manuscript was written by the author,  who takes full responsibility for any errors.

\appendix

\sectiono{Definition of the world-sheet operator $t$} \label{sb}

\newcommand{\Ph}{\widehat\Phi}

In flat space-time background the world-sheet fields corresponding to the target space coordinates
are not good conformal fields, but their derivatives are. Nevertheless we can define
\be
X = -i\, {\p\over \p k} \, e^{ikX}\bigg|_{k=0}\, ,
\ee
and compute correlation functions of $X$ using this representation. In this appendix we shall
give a similar definition for the field $t(z,\bar z)$ for the $SL(2,R)$ WZW model. 

The primaries $\Phi_j(x,\bar x|z,\bar z)$ 
of this theory can be labelled by quantum numbers $j, x, \bar x$ where 
$x$, $\bar x$
are the coordinates on the conformal boundary of the Euclidean 
$AdS_3$\cite{9712256,9712258}. We shall work in global $AdS_3$ before the periodic identification of the $t$
coordinate. We define,
\be
\Phi_j(\infty| z,\bar z) = \lim_{x\to\infty, \bar x\to\infty} (x^{-2j}\,  \bar x^{-2j}) \Phi_j(x,\bar x; z,\bar z)\, .
\ee
In terms of this, we define
\begin{equation}\label{main}
t(z,\bar z)=\lim_{j\to 0}\frac14\,\partial_j\Big[\Phi_j(0|z,\bar z)-\Phi_j(\infty|z,\bar z)\Big]\, .
\end{equation}
Here $\Phi_j(x|z)$ are the affine primaries and $x=0,\infty$ are the two boundary points fixed by the $J^3$ generator. The $j\to 0$ limit has to be defined via analytic continuation.

We can check this by taking the classical limit in which $\Phi_j$ becomes a classical function of
the coordinates of $AdS_3$. We have\cite{9712258}
\be
\Phi_j(x,\bar x) = (e^{-\varpi}+(\gamma-x)(\bar\gamma-\bar x) e^\varpi)^{2j}\, ,
\ee
where $\gamma,\bar\gamma,\varpi$ are related to the coordinates $t,\rho,\varphi$ appearing in
\refb{e2.1} via
\be
e^\varpi = e^{-t}\cosh\rho, \qquad \gamma = e^{t+ i\vp} \tanh\rho,
\qquad \bar\gamma = e^{t- i\vp} \tanh\rho\, .
\ee
Then we have
\ben
\Phi_j(0) &=& (e^{-\varpi} + e^{\varpi}\, \bar\gamma\gamma)^{2j} = (e^t / \cosh\rho+ e^{t}\tanh^2\rho\,
\cosh\rho)^{2j},
\nonumber \\
\Phi_j(\infty) &=& \lim_{x\to\infty, \bar x\to \infty} x^{-2j} \bar x^{-2j} (x\bar x e^{-t}\cosh\rho)^{2j}
= e^{-2jt}\cosh^{2j}(\rho)\, .
\een
Then
\be
\frac14\,\partial_j\Big[\Phi_j(0|z,\bar z)-\Phi_j(\infty|z,\bar z)\Big]_{j=0}=\frac12\,
\left[\ln (e^t / \cosh\rho+ e^{t}\tanh^2\rho \cosh\rho) - \ln (e^{-t}\cosh\rho)
\right]=t\, .
\ee
This is the desired relation.

\sectiono{Some torus correlators} \label{sa}

In this appendix we shall compute the torus two point function of $\p t$ and $\bar\p t$ with
various currents.

First consider the correlator $J^3(z) \bar \p t(w)$. This is a holomorphic function of $z$ away
from $z=w$ and it follows from \refb{e3.14e} that it has no singularity as $z\to w$ except 
possibly delta function contact terms. This shows that the correlator must be independent of
$z$ away from $z=w$. Translation invariance on the torus then implies that it is also 
independent of $w$. Hence
we can write
\be\label{edelta}
\langle J^3(z) \bar \p t(w) \rangle  = C + {\pi\over 2} \delta^{(2)}(z-w) \langle 1\rangle\, ,
\ee
where the term proportional to $\delta^{(2)}(z-w)$ is computed using the operator product
expansion \refb{e3.14e}. We now integrate both sides over the torus. 
Using the fact that the
left hand side is a total derivative we can express this as a boundary contribution. Since
$t(w+\tau)=t(w)+2\pi\uptau_2$, we can use \refb{e3.7a} to conclude that integral of
the left hand side gives
$
i\pi\uptau_2\, \langle J^3(w)\rangle
$.  
On the other hand the integral of the right hand side is given by  $C\tau_2 +{\pi \over 2}$.
Hence we have 
\be
i\pi\uptau_2\, \langle J^3(w)\rangle = C\tau_2 +{\pi \over 2} \langle 1\rangle\, .
\ee 
This gives, away from $z=w$,
\be
\langle J^3(z) \bar \p t(w) \rangle = C = {i\pi\uptau_2\over \tau_2}\, \langle J^3(w)\rangle
- {\pi\over 2\, \tau_2} \langle 1\rangle\, .
\ee
For $K^3(z) \bar\p t(w)$ the analysis is similar except that we do not have the 
delta function term in \refb{edelta}. Hence we get
\be
\langle K^3(z) \bar \p t(w) \rangle = {i\pi\uptau_2\over \tau_2}\, \langle K^3(w)\rangle\, .
\ee
Similar analysis gives
\be
\langle \bar J^3(z')  \p t(w) \rangle = - {i\pi\uptau_2\over \tau_2}\, \langle \bar J^3(w)\rangle
- {\pi\over 2\, \tau_2}\langle 1\rangle\, ,
\ee
and 
\be
\langle \bar K^3(z') \p t(w) \rangle = -{i\pi\uptau_2\over \tau_2}\, \langle \bar K^3(w)\rangle\, .
\ee


\begin{thebibliography}{99}

\bibitem{9711200}
J.~M.~Maldacena,
``The Large $N$ limit of superconformal field theories and supergravity,''
Adv. Theor. Math. Phys. \textbf{2} (1998), 231-252
doi:10.4310/ATMP.1998.v2.n2.a1
[arXiv:hep-th/9711200 [hep-th]].

\bibitem{9806194}
A.~Giveon, D.~Kutasov and N.~Seiberg,
``Comments on string theory on AdS(3),''
Adv. Theor. Math. Phys. \textbf{2} (1998), 733-782
doi:10.4310/ATMP.1998.v2.n4.a3
[arXiv:hep-th/9806194 [hep-th]].


\bibitem{9903219}
D.~Kutasov and N.~Seiberg,
``More comments on string theory on AdS(3),''
JHEP \textbf{04} (1999), 008
doi:10.1088/1126-6708/1999/04/008
[arXiv:hep-th/9903219 [hep-th]].

\bibitem{0001053}
J.~M.~Maldacena and H.~Ooguri,
``Strings in AdS(3) and SL(2,R) WZW model 1.: The Spectrum,''
J. Math. Phys. \textbf{42} (2001), 2929-2960
doi:10.1063/1.1377273
[arXiv:hep-th/0001053 [hep-th]].

\bibitem{0005183}
J.~M.~Maldacena, H.~Ooguri and J.~Son,
``Strings in AdS(3) and the SL(2,R) WZW model. Part 2. Euclidean black hole,''
J. Math. Phys. \textbf{42} (2001), 2961-2977
doi:10.1063/1.1377039
[arXiv:hep-th/0005183 [hep-th]].

\bibitem{0111180}
J.~M.~Maldacena and H.~Ooguri,
``Strings in AdS(3) and the SL(2,R) WZW model. Part 3. Correlation functions,''
Phys. Rev. D \textbf{65} (2002), 106006
doi:10.1103/PhysRevD.65.106006
[arXiv:hep-th/0111180 [hep-th]].

\bibitem{1803.04423}
M.~R.~Gaberdiel and R.~Gopakumar,
``Tensionless string spectra on AdS$_{3}$,''
JHEP \textbf{05} (2018), 085
doi:10.1007/JHEP05(2018)085
[arXiv:1803.04423 [hep-th]].

\bibitem{1812.01007}
L.~Eberhardt, M.~R.~Gaberdiel and R.~Gopakumar,
``The Worldsheet Dual of the Symmetric Product CFT,''
JHEP \textbf{04} (2019), 103
doi:10.1007/JHEP04(2019)103
[arXiv:1812.01007 [hep-th]].

\bibitem{1911.00378}
L.~Eberhardt, M.~R.~Gaberdiel and R.~Gopakumar,
``Deriving the AdS$_{3}$/CFT$_{2}$ correspondence,''
JHEP \textbf{02} (2020), 136
doi:10.1007/JHEP02(2020)136
[arXiv:1911.00378 [hep-th]].

\bibitem{2009.11306}
A.~Dei, M.~R.~Gaberdiel, R.~Gopakumar and B.~Knighton,
``Free field world-sheet correlators for ${\rm AdS}_3$,''
JHEP \textbf{02} (2021), 081
doi:10.1007/JHEP02(2021)081
[arXiv:2009.11306 [hep-th]].

\bibitem{2011.10038}
M.~R.~Gaberdiel, R.~Gopakumar, B.~Knighton and P.~Maity,
``From symmetric product CFTs to AdS$_{3}$,''
JHEP \textbf{05} (2021), 073
doi:10.1007/JHEP05(2021)073
[arXiv:2011.10038 [hep-th]].

\bibitem{2306.16280}
N.~M.~McStay and R.~A.~Reid-Edwards,
``Symmetries and covering maps for the minimal tension string on 
AdS$_{3}\times$S$^{3}\times$T$^{4}$,''
JHEP \textbf{04} (2024), 048
doi:10.1007/JHEP04(2024)048
[arXiv:2306.16280 [hep-th]].

\bibitem{2610.20696}
L.~Eberhardt, M.~R.~Gaberdiel, S.~Murthy and M.~Rangamani,
``Worldsheet string theory on $\mathrm{AdS}_3 \times \mathbf{S}^3 \times \mathrm{K3}$,''
[arXiv:2610.20696 [hep-th]].

\bibitem{2408.14567}
C.~Ferko, S.~Murthy and M.~Rangamani,
``Strings in AdS$_{3}$: one-loop partition function and near-extremal BTZ thermodynamics,''
JHEP \textbf{05} (2025), 010
doi:10.1007/JHEP05(2025)010
[arXiv:2408.14567 [hep-th]].

\bibitem{2505.14380}
S.~Murthy and M.~Rangamani,
``Strings and near-extremal black holes in theories with large $ \mathcal{N} $ = 4 superconformal symmetry,''
JHEP \textbf{04} (2026), 142
doi:10.1007/JHEP04(2026)142
[arXiv:2505.14380 [hep-th]].

\bibitem{1501.00988}
A.~Sen,
``Gauge Invariant 1PI Effective Superstring Field Theory: Inclusion of the Ramond Sector,''
JHEP \textbf{08} (2015), 025
doi:10.1007/JHEP08(2015)025
[arXiv:1501.00988 [hep-th]].

\bibitem{1508.05387}
A.~Sen,
``BV Master Action for Heterotic and Type II String Field Theories,''
JHEP \textbf{02} (2016), 087
doi:10.1007/JHEP02(2016)087
[arXiv:1508.05387 [hep-th]].

\bibitem{1703.06410}
C.~de Lacroix, H.~Erbin, S.~P.~Kashyap, A.~Sen and M.~Verma,
``Closed Superstring Field Theory and its Applications,''
Int. J. Mod. Phys. A \textbf{32} (2017) no.28n29, 1730021
doi:10.1142/S0217751X17300216
[arXiv:1703.06410 [hep-th]].

\bibitem{2607.12031}
A.~Sen,
``A New Term in Type II Effective Action,''
[arXiv:2607.12031 [hep-th]].

\bibitem{2405.19421}
A.~Sen and B.~Zwiebach,
``String Field Theory: A Review,''
[arXiv:2405.19421 [hep-th]].

\bibitem{Polchinski:1985zf}
J.~Polchinski,
``Evaluation of the One Loop String Path Integral,''
Commun. Math. Phys. \textbf{104} (1986), 37
doi:10.1007/BF01210791

\bibitem{9712256}
J.~Teschner,
``On structure constants and fusion rules in the SL(2,C) / SU(2) WZNW model,''
Nucl. Phys. B \textbf{546} (1999), 390-422
doi:10.1016/S0550-3213(99)00072-3
[arXiv:hep-th/9712256 [hep-th]].


\bibitem{9712258}
J.~Teschner,
``The Minisuperspace limit of the sl(2,C) / SU(2) WZNW model,''
Nucl. Phys. B \textbf{546} (1999), 369-389
doi:10.1016/S0550-3213(99)00071-1
[arXiv:hep-th/9712258 [hep-th]].



\end{thebibliography}
\end{document}